\input harvmac
\newcount\figno

\def\encadremath#1{\vbox{\hrule\hbox{\vrule\kern8pt\vbox{\kern8pt
\hbox{$\displaystyle #1$}\kern8pt}
\kern8pt\vrule}\hrule}}

\overfullrule=0pt

%
\def\tilde{\widetilde}
\def\bar{\overline}
\def\Z{{\bf Z}}

\def\S{{\bf S}}
\def\R{{\bf R}}
\def\npb#1#2#3{{\it Nucl.\ Phys.} {\bf B#1} (19#2) #3}
\def\plb#1#2#3{{\it Phys.\ Lett.} {\bf B#1} (19#2) #3}

\font\zfont = cmss10 

\def\bigone{\hbox{1\kern -.23em {\rm l}}}
\def\ZZ{\hbox{\zfont Z\kern-.4emZ}}

\def\hat{\widehat}
\def\frac#1#2{{#1\over #2}}
\def\CC{{\cal C}}
\Title{\vbox{\baselineskip12pt
\hbox{hep-th/9603142}
\hbox{IASSNS-HEP-96/17}
\hbox{PUPT-1597}}}
{\vbox{\centerline{ELEVEN-DIMENSIONAL SUPERGRAVITY}
\bigskip
\centerline{ ON A MANIFOLD WITH BOUNDARY}}}
\smallskip
\centerline{Petr Ho\v rava\footnote{$^\ast$}{horava@puhep1.princeton.edu.  
Research supported in part by NSF Grant PHY90-21984.}}
\smallskip
\centerline{\it Joseph Henry Laboratories, Princeton University}
\centerline{\it Jadwin Hall, Princeton, NJ 08544, USA}
\smallskip
\centerline{and}
\smallskip
\centerline{Edward Witten\footnote{$^\star$}{witten@sns.ias.edu.  
Research supported in part by NSF Grant PHY95-13835.}}
\smallskip
\centerline{\it School of Natural Sciences, Institute for Advanced Study}
\centerline{\it Olden Lane, Princeton, NJ 08540, USA}\bigskip
\medskip

In this paper, we present a systematic analysis of eleven-dimensional 
supergravity on a manifold with boundary, which is believed to be relevant to 
the strong coupling limit of the $E_8\times E_8$ heterotic string.  Gauge and 
gravitational anomalies enter at a very early stage, and require a refinement 
of the standard Green-Schwarz mechanism for their cancellation.  This 
uniquely determines the gauge group to be a copy of $E_8$ for each boundary 
component, fixes the gauge coupling constant in terms of the gravitational 
constant, and leads to several striking new tests of the hypothesis that 
there is a consistent quantum $M$-theory with eleven-dimensional supergravity 
as its low energy limit.

\Date{March, 1996}
\nref\hw{P. Ho\v rava and E. Witten, ``Heterotic and Type I String Dynamics 
from Eleven Dimensions,'' \npb{460}{96}{506}, hep-th/9510209.}
\nref\van{E. Bergshoeff, M. de~Roo, B. de~Wit and P. van~Nieuwenhuisen, 
``Ten-Dimensional Maxwell-Einstein Supergravity, its Currents and the Issue 
of its Auxiliary Fields,'' \npb{195}{82}{97}.}
\nref\chap{G. Chapline and N.S. Manton, ``Unification of Yang-Mills Theory 
and Supergravity in Ten Dimensions,'' \plb{120}{83}{105}.}
\nref\duffliu{M.J. Duff, J.T. Liu, and R. Minasian, ``Eleven-Dimensional 
Origin of String/String Duality: A One Loop Test,'' \npb{452}{95}{261}, 
hep-th/9506126.}
\nref\blum{J.A. Harvey and J. Blum, unpublished.}
\nref\witten{E. Witten, ``Five-Branes and $M$-Theory on an Orbifold,'' 
{\it Nucl.\ Phys.} {\bf B463} (1996), hep-th/9512219.}
\nref\vafa{C. Vafa and E. Witten, ``A One-Loop Test of String Duality,'' 
\npb{447}{95}{261}, hep-th/9505053.}
\nref\julia{E. Cremmer, B. Julia, and J. Scherk, ``Supergravity Theory in 
11 Dimensions,'' \plb{76}{78}{409}.}
\newsec{Introduction}

In a previous paper \hw, we proposed that the strong coupling limit of the 
ten-dimensional $E_8\times E_8$ heterotic string is eleven-dimensional 
$M$-theory compactified on $\R^{10}\times \S^1/\Z_2=\R^{10}\times I$ ($I$ is 
the unit interval), with the gauge fields entering via ten-dimensional 
vector multiplets that propagate on the boundary of space-time.  This 
implies in particular that there must exist a supersymmetric coupling of 
ten-dimensional vector multiplets on the boundary of an eleven-manifold 
to the eleven-dimensional supergravity multiplet propagating in the bulk.  
The purpose of the present paper is to explore this coupling.  

In doing so, one comes quickly to a puzzle.  The supergravity action in bulk 
is  
\eqn\grabulk{-{1\over 2\kappa^2}\int_{M^{11}} d^{11}x\sqrt g\, R + \dots,}
with $M^{11}$ being the eleven-dimensional space-time, and ``$\dots$'' being 
the terms involving fermions and the bosonic three-form field.  The 
supergauge action on the boundary is 
\eqn\gaugebulk{-{1\over 4\lambda^2}\int_{M^{10}}d^{10}x\sqrt g\,\tr\,F^2
+\dots,}
where $M^{10}$ is the boundary (or a component of the boundary) of $M^{11}$, 
and $F$ is the field strength of the gauge fields that propagate 
on $M^{10}$. (For $E_8\times E_8$, ``$\tr $'' is as usual $1/30$ of the 
trace $\Tr$ in the adjoint representation.)  
In the above formulas, $\kappa$ and $\lambda$ are the 
gravitational and gauge coupling constants.  From those constants one 
can make a dimensionless number $\eta=\lambda^6/\kappa^4$.  
The question is what determines the value of $\eta$.  Note that there is 
no dilaton or other scalar whose expectation value controls the value of 
$\eta$.  In fact, there is no scalar field at all in the theory, 
propagating either in bulk or on the boundary; in going to strong coupling, 
the dilaton of the perturbative heterotic string is reinterpreted as 
the radius of $\S^1/\Z_2$.  

Since string theory has no adjustable parameter corresponding to $\eta$, 
the strong coupling limit of the $E_8\times E_8$ heterotic string, if it does 
have the eleven-dimensional interpretation proposed in our previous paper, 
must give a definite value for $\eta$.  In fact, we will argue in this 
paper that by looking more precisely at gravitational and gauge anomalies 
(which were already used in the previous paper), one can determine $\eta$.  
We get 
\eqn\deteta{\eta =128\pi^5 ,}
or equivalently 
\eqn\jeteta{\lambda^2=2\pi(4\pi\kappa^2)^{2/3}.}   

In the remainder of this introduction, we sketch the argument that will be 
used to determine $\eta$, and also sketch the other main qualitative results 
of this paper.  The reason for presenting such a detailed sketch first is 
that the supergravity calculation that occupies the remainder of the paper is 
unavoidably rather complicated.

We recall that anomalies in ten dimensions are described by a formal 
twelve-form $I_{12}(R,F_1,F_2)$ that is a sixth order homogeneous polynomial 
in the Riemann tensor $R$ and the field strengths $F_1$ and $F_2$ in the two 
$E_8$'s.  It has the general form 
\eqn\genform{I_{12}(R,F_1,F_2) = A(R) +B(R,F_1) +B(R,F_2),}
where $A(R)$ is the contribution of the supergravity multiplet, and 
$B(R,F_i)$, for $i=1,2$, is the contribution of the gluinos of the $i^{th}$ 
$E_8$.  In \hw, we introduced 
\eqn\penform{\hat I_{12}(R,F)={1\over 2}A(R) +B(R,F),}
so that
\eqn\henform{I_{12}(R,F_1,F_2)=\hat I_{12}(R,F_1) +\hat I_{12}(R,F_2).}
The idea here is  that from the eleven-dimensional point of view, the gauge 
and gravitational anomaly is localized on the boundary of space-time, and 
the two terms on the right of \henform\ are the contributions of the two 
components of the boundary of $\R^{10}\times I$.  Of the two $E_8$'s, the 
one propagating on a given boundary component is naturally the only one that 
contributes to the anomaly form of that component.  

Anomaly cancellation for the perturbative heterotic string involves a 
factorization 
\eqn\facform{I_{12}= I_4 I_8,}
where $I_4=\tr\,R^2-\tr\,F_1^2-\tr\,F_2^2$ and $I_8$ is an eight-form given 
by a lengthy quartic polynomial in $R$ and the $F_i$.  As was explained in 
\hw, $\hat I_{12}$ has an analogous factorization 
\eqn\gacform{\hat I_{12} =\hat I_4 \hat I_8,}
with 
\eqn\lenform{\eqalign{\hat I_4(R,F)=&{1\over 2}\tr\,R^2-\tr\,F^2 \cr
                      \hat I_8(R,F)=&  -{1\over 4}\hat I_4(R,F)^2
+\left(-{1\over 8}\tr\, R^4+{1\over 32} (\tr\, R^2)^2\right).\cr}}
(This way of writing the formula for $\hat I_8$, which was also noticed by 
M.~Duff and R.~Minasian, has a rationale that will become clear in section 
three.)  It was proposed in \hw\ that this factorization of $\hat I_{12}$ 
would permit an extension of the Green-Schwarz anomaly cancellation 
mechanism to $M$-theory on eleven-dimensional manifolds with boundary.  

The Green-Schwarz mechanism in ten dimensions depends on the existence in 
string theory of a two-form field $B$ whose gauge-invariant field strength 
$H$ obeys 
\eqn\aka{ dH = I_4.}
Such an equation (with only the $\tr\,F^2$ term in $I_4$) holds even in the 
minimal ten-dimensional supergravity \refs{\van,\chap}.  In addition, there 
are ``Green-Schwarz interaction terms,'' present in the string theory but 
{\it not\/} in the minimal low energy supergravity theory, of the form 
\eqn\baka{\Delta L = \int B\wedge I_8.} 
The combination of \aka\ and \baka\ gives a classical theory that 
is {\it not\/} gauge invariant, with an anomaly constructed from the 
twelve-form $I_{12}= I_4I_8$.  The minimal classical supergravity theory is 
gauge invariant because the anomalous fermion loops and the Green-Schwarz 
terms are both absent, and the string theory is gauge 
invariant because they are both present and the anomalies cancel 
between them.  

Now let us discuss how the story will work in eleven dimensions.  In doing 
so, and in most of this paper,  we will use an orbifold approach in which we 
work on an  eleven-manifold $M^{11}$ with a $\Z_2$ symmetry whose fixed 
points are of codimension one; alternatively, one can take the quotient and 
work on the manifold-with-boundary $X=M^{11}/\Z_2$, whose boundary points 
are the $\Z_2$ fixed points in $M^{11}$.  In general, the formulation 
in terms of a manifold  with boundary is convenient intuitively, 
and the orbifold formulation is convenient for calculation.  

Rather than a two-form $B$, the eleven-dimensional supergravity multiplet 
has a three-form field $C$ (denoted by $A^{(3)}$ in our previous paper \hw), 
whose field strength is a four-form $G$.%
\foot{Following conventions in \van\ which have become standard 
in eleven-dimensional supergravity, we define $G_{IJKL}=\partial_IC_{JKL} 
\pm  23 \,\,{\rm terms}$, though the normalization is somewhat unusual. 
We also define $dG_{IJKLM}=\partial_IG_{JKLM}+{\rm cyclic\,\,permutations\,\,
of}\,\,IJKLM$.}
In the absence of boundaries, $G$ obeys the usual Bianchi identity $dG=0$.  
The analog of \aka\ will have to be a contribution to $dG$ supported at the 
$\Z_2$ fixed points.  As $dG$ is a five-form, we will have to promote the 
four-form $\hat I_4$ to a five-form supported on the fixed point set, so 
that it can appear as a correction to the Bianchi identity.  
To write such a five-form, one supposes that the fixed point set is defined 
locally by an equation $x^{11}=0$, and one multiplies by the closed 
one-form $ \delta(x^{11}) dx^{11} $ to promote the four-form $\hat I$ 
to a five-form.  Thus, the eleven-dimensional  analog of the ten-dimensional 
 equation $dH=I_4$ will be an equation $dG = c \delta(x^{11}) dx^{11} \hat 
I_4$, with some constant $c$.  In section two, we will determine the precise 
equation to be 
\eqn\daka{dG_{11\,IJKL}=-{3\sqrt 2\over 2\pi}\left(\kappa\over 4\pi
\right)^{2/3} \delta(x^{11})\left(\tr\,F_{[IJ}F_{KL]} -{1\over 2}\tr\,
R_{[IJ}R_{KL]}\right).}
Here $F$ is of course the field strength of the gauge field propagating 
at $x^{11}=0$, and 
$\tr\,F_{[IJ}F_{KL]}=(1/24)\tr\, F_{IJ}F_{KL}\pm\;{\rm permutations}$.  
Actually, in section two, we will see directly only the $\tr\,F\wedge F$ term 
in \daka; the $\tr\,R\wedge R$ term is a sort of higher order 
correction that we infer because it is needed for anomaly cancellation.  
(Analogously, in ten dimensions, the $\tr\,F\wedge F$ term is required 
by supersymmetry, and the $\tr\,R\wedge R$ term is an $O(\alpha')$ stringy 
correction needed for anomaly cancellation.)  

At this stage the question is, what are the Green-Schwarz terms? 
In the familiar ten-dimensional story, because the Green-Schwarz terms 
are unconstrained by supersymmetry, the Green-Schwarz mechanism makes 
no general prediction (independent of anomalies or a detailed string 
model) about what $I_8$ should be.  In eleven dimensions, the story will 
be quite different because the terms analogous to the Green-Schwarz 
terms are independently known.  One of these terms is simply 
the $\int C\wedge G\wedge G$ interaction of eleven-dimensional 
supergravity.  This term, discovered when the model was first 
constructed \julia  , has always 
seemed enigmatic because the rationale behind its apparently ``topological'' 
nature was not clear. We feel that the role of this term in canceling 
anomalies -- we explain in section three how $\int C\wedge G\wedge G$ 
comes to play the role of a Green-Schwarz term -- removes some of 
the enigma.  

The $\int C\wedge G \wedge G$ is the only ``Green-Schwarz'' interaction 
involved in canceling gauge anomalies, but to cancel also the gravitational 
anomalies requires an additional interaction.  This is an eleven-dimensional 
interaction 
\eqn\haka{\int_{M^{11}} C\wedge X_8(R),}
with $X_8(R)$ an eight-form constructed as a quartic polynomial in the 
Riemann tensor.  This interaction is known in two ways.  (1) Upon dimensional 
reduction on $\S^1$, it turns into a $B\wedge X_8$ interaction which can be 
computed as a one-loop effect in Type IIA superstring theory \vafa .  
The one loop calculation is exact since a dilaton dependence of the 
$B\wedge X_8$ coupling would spoil gauge invariance; because it is exact, 
it can be extrapolated to eleven dimensions and implies the existence of the 
interaction written in \haka.  (2) Alternatively, this coupling is needed to 
cancel one-loop anomalies on the five-brane world-volume and thus permit the 
existence of five-branes in the theory \refs{\duffliu - \witten}.  Happily, 
the two methods agree, with
\eqn\waka{X_8=-{1\over 8}\tr\,R^4+{1\over 32} (\tr\, R^2)^2 .}
As we will see in section three, it is no coincidence that the combination 
of $\tr\,R^4$ and $(\tr\,R^2)^2$ that appears here also entered in \lenform.  

The fact that the terms analogous to Green-Schwarz terms are known 
independently of any discussion of space-time anomalies means that we get 
an {\it a priori\/} prediction for $\hat I_8$.  (We have no {\it a priori\/} 
prediction of $\hat I_4$, as the coefficients in \daka\ will essentially be 
adjusted to make anomaly cancellation possible.)  We regard the success of 
this prediction as a compelling confirmation that eleven-dimensional 
supergravity on a manifold with boundary is indeed related to ten-dimensional 
$E_8\times E_8$ heterotic string theory as proposed in \hw.  

\bigskip\noindent{\it Classical and Quantum Consistency}

The details that we have just explained of how anomaly cancellation works in 
eleven dimensions have other implications for the structure of the theory.

The fact that, once one works on a manifold with boundary, some of the 
Green-Schwarz terms are present in the minimal supergravity Lagrangian means 
that the classical Lagrangian, including the vector supermultiplets on the 
boundary, is not gauge invariant.  Thus, the theory with the supergravity 
multiplet in bulk and the vector multiplets on the boundary is only 
consistent as a quantum theory.  The situation is rather different 
from perturbative string theory, where since the Green-Schwarz terms arise 
at the one loop level, one has gauge invariance either classically 
(leaving out the anomalous chiral fermion loop diagrams and the effects 
of the Green-Schwarz terms) or quantum mechanically (including both 
of these).  

The relation $\lambda^2\sim \kappa^{4/3}$ between the gauge and gravitational 
couplings sheds a further light on this.  It means that the gauge kinetic 
energy, of order $1/\lambda^2$, is a higher order correction, of relative 
order $\kappa^{2/3}$ compared to the gravitational action, which is of order 
$1/\kappa^2$.  If one wants a fully consistent classical theory, one must 
ignore the gauge fields completely.  Once one tries to include the supergauge 
multiplet, gauge invariance will fail classically (in relative order 
$\kappa^2$), and quantum anomalies are needed to compensate for this failure.

Since the classical theory with the gauge fields is not going to be fully 
consistent, one has to expect peculiarities in constructing it.  In our 
analysis in section four, we certainly find such peculiarities.  We will 
organize our discussion of the boundary interactions as an expansion in 
powers of $\kappa^{2/3}$.  In order $\kappa^{2/3}$, things go smoothly, 
though the calculations are rather involved, roughly as in standard 
supergravity theories.  Some novelties arise in order $\kappa^{4/3}$.  In 
verifying invariance of the Lagrangian in that order, one has to cancel terms 
that are formally proportional to $\delta(0)$.  The cancellation also 
involves adding to the Lagrangian new interactions (of relative order 
$\kappa^{4/3}$) proportional to $\delta(0)$.  We interpret the occurrence of 
$\delta(0)$ terms in the Lagrangian and the supersymmetry variations of 
fields as a symptom of attempting to treat in classical supergravity what 
really should be treated in quantum $M$-theory.  In a proper quantum 
$M$-theory treatment, there would presumably be a built-in cutoff that would 
replace $\delta(0)$ by a finite constant times $\kappa^{-2/9}$.  For 
instance, the cutoff might involve having the gauge fields propagate in a 
boundary layer, with a thickness of order $\kappa^{2/9}$, and not precisely 
on the boundary of space-time.  

Though the $\delta(0)$ terms formally cancel in order $\kappa^{4/3}$, one 
must expect further difficulties in higher order since, without knowing 
the correct way to cut off the linear divergence that gave the $\delta(0)$ 
terms in order $\kappa^{4/3}$, there is some uncertainty in the determination 
of the correct structure in that order.  One must suppose, by analogy with 
many other problems in physics, that underneath the cancellation of the 
linear divergences there might be a finite remainder, which could be 
extracted if one understood the correct cutoff.  Without understanding 
the finite remainder, one should expect difficulty in proceeding to the next 
order.  

In any event, one of the things that happens in the next order -- relative 
order $\kappa^2$ -- has already been explained.  One runs into a failure of 
classical gauge invariance which must be canceled by quantum one-loop 
anomalies (which are also of relative order $\kappa^2$).  It is hard to 
believe that the classical discussion can usefully be continued to higher 
order, once the classical gauge invariance has failed and one has begun to 
run into conventional quantum loops.  An attempt to continue the classical 
discussion  would almost undoubtedly soon run into higher order divergences 
than the $\delta(0)$ that we described two paragraphs ago; for instance, one 
would very likely find $\delta(0)$ terms in the supergravity transformation 
laws and $\delta(0)^2$ terms in the Lagrangian.

Despite the infinities that arise in the construction, we hope and expect 
that the analysis of the anomalies is reliable.  This should be analogous to 
the fact that anomalous loop diagrams can be reliably computed even in 
unrenormalizable effective theories, because the anomalies can be construed 
as an infrared effect and are independent of what cutoff one introduces.

\bigskip\noindent{\it Summary}

To summarize, then, the lessons from our investigation, we will find that 
anomaly cancellation of the ten-dimensional heterotic string has an elegant 
eleven-dimensional interpretation that sheds light on properties of the 
anomaly twelve-form that were not needed before.  This sharpens the 
eleven-dimensional interpretation of the strongly coupled $E_8\times E_8$ 
heterotic string, fixing an otherwise unknown dimensionless parameter and 
adding to our confidence that the eleven-dimensional description is correct.  
The gauge anomalies that arise in the classical discussion also give an 
indication -- and not the only one -- that the theory only really makes 
sense at the quantum level.  

\newsec{Correction to the Bianchi Identity}

Our eleven-dimensional conventions are as in \van.  We work with Lorentz 
signature $-++\dots +$.  Vector indices will be written as $I,J,K$, and 
spinor indices as $\alpha,\beta,\gamma$.   We introduce a frame field 
$e_I{}^m$ with the metric being $g_{IJ}=\eta_{mn}e_I{}^me_J{}^n$.  The gamma 
matrices are $32\times 32$ real matrices obeying $\{\Gamma_I,\Gamma_J\}=
2g_{IJ}$.  One also defines $\Gamma^{I_1I_2\dots I_n}=\Gamma^{[I_1}\dots
\Gamma^{I_n]}\equiv (1/n!)\Gamma^{I_1}\Gamma^{I_2}\dots\Gamma^{I_n}\pm{\rm 
permutations}$.  Spinor indices are raised and lowered with a real 
antisymmetric tensor $\CC$ obeying $\CC_{\alpha\beta}=-\CC_{\beta\alpha}$, 
$\CC^{\alpha\beta}\CC_{\beta\gamma}=\delta^\alpha_\gamma$.  
In particular, by lowering an index in the gamma matrix ${\Gamma_I^\alpha
}_\beta$ one gets a symmetric tensor $\Gamma_{I\alpha\beta}=\Gamma_{I\beta
\alpha}$.  All spinors will be Majorana spinors; the symbol $\bar\psi_\alpha$ 
is simply defined by $\bar\psi_\alpha=\CC_{\alpha\beta}\psi^\beta$.  

The supergravity multiplet consists of the metric $g$, the gravitino 
$\psi_{I\alpha}$, and a three-form $C$ (with field strength $G$, normalized 
as in a previous footnote).  The supergravity Lagrangian, up to terms 
quartic in the gravitino (which we will not need), is \julia\  
\eqn\suglag{\eqalign{L_S&={1\over \kappa^2}\int_{M^{11}}d^{11}x\sqrt g
\left(-{1\over 2}R -{1\over 2}\bar\psi_I\Gamma^{IJK}D_J\psi_K
-{1\over 48}G_{IJKL}G^{IJKL}\right. \cr &\qquad\qquad\qquad\qquad
{}-{\sqrt 2\over 192}\left(\bar\psi_I\Gamma^{IJKLMN}\psi_N 
 +12\bar \psi^J\Gamma^{KL}\psi^M\right)G_{JKLM}\cr 
&\qquad\qquad\qquad\qquad\qquad\qquad\left.{}-{\sqrt 2\over 3456}
\epsilon^{I_1I_2\dots I_{11}}C_{I_1I_2I_3}G_{I_4\dots I_7}G_{I_8\dots I_{11}}
\right).\cr}}
We work in $1.5$ order formalism:  the spin connection $\Omega$ is formally 
regarded as an independent variable, and eventually set equal to the 
solution of the bulk equations of motion.  The Riemann tensor is the field 
strength constructed from $\Omega$.

The transformation laws of local supersymmetry read 
\eqn\tranlaws{\eqalign{\delta e_I{}^m & = {1\over 2}\bar\eta\Gamma^m\psi_I \cr
                       \delta C_{IJK} & = -{\sqrt 2\over 8}\bar\eta
                         \Gamma_{[IJ}\psi_{K]} \cr
                        \delta\psi_I& =D_I\eta+{\sqrt 2\over 288} 
\left(\Gamma_I{}^{JKLM} -8\delta_I^J\Gamma^{KLM}\right)\eta G_{JKLM}+\dots.
\cr}}
(The $\dots$ are three fermi terms in the transformation law of $\psi$, often 
absorbed in a definition of ``supercovariant'' objects; we will not need 
them.)  

We suppose that there is a $\Z_2$ symmetry acting on $M^{11}$, with 
codimension one fixed points.  We let $M^{10}$ be a component of the fixed 
point set; we will study the physics near $M^{10}$.  We suppose that the 
fields are required to be invariant under the $\Z_2$; this means that we 
could pass to the manifold-with-boundary $X=M^{11}/\Z_2$ (with boundary 
$M^{10}$), but that will not be particularly convenient.  If $M^{10}$ is 
defined locally  by an equation $x^{11}=0$, $x^{11}$ being one of the 
coordinates (and the $\Z_2$ acting by $x^{11}\to -x^{11}$), then (with an 
appropriate lifting of the $\Z_2$ action to spinors and the three-form) 
the supersymmetries that commute with the $\Z_2$ action are generated 
by spinor fields $\eta$ on $M^{11}$ that obey 
\eqn\plunk{\Gamma_{11}\eta= \eta \,\,\,{\rm at}\,\, x^{11} =0.}
$\Z_2$ invariance of the gravitino means that 
\eqn\ztwo{\eqalign{\Gamma_{11} \psi_A & = \psi_A, \qquad A=1,\dots, 10 \cr
                   \Gamma_{11}\psi_{11} & = -\psi_{11} .\cr}}
As in \ztwo, we will use $A,B,C,D=1,\dots, 10$ for indices tangent 
to $M^{10}$.  For the three-form $C$, because it is odd under parity 
(this follows from the $CGG$ interaction in \suglag), $\Z_2$ invariance means 
that $C_{BCD}=0$ at $x^{11}=0$.  A gauge-invariant statement that follows 
from this is that \eqn\yutwo{G_{ABCD} = 0 \,\,\,\,{\rm at }\,\,x^{11}=0,} 
or in other words, the pull-back of the differential form $G$ to $M^{10}$ 
vanishes.  We will eventually find a sort of modification of this statement 
in order $\kappa^{2/3}$.    

The vector supermultiplets, which propagate on $M^{10}$, consist of the $E_8$ 
gauge field $A$ (with field strength $F_{CD}=\partial_CA_D-\partial_DA_C+
[A_C,A_D]$) and fermions (gluinos) $\chi$ in the adjoint representation, 
obeying $\Gamma_{11}\chi=\chi$.  The minimal Yang-Mills Lagrangian is 
\eqn\ymlag{L_{YM}=-{1\over \lambda^2}\int_{M^{10}}d^{10}x\sqrt g\,\,
\tr\left({1\over 4}F_{AB}F^{AB}+{1\over 2}\bar\chi \Gamma^AD_A\chi\right).}  
(Here $d^{10}x\sqrt g$ is understood as the Riemannian measure of $M^{10}$, 
using the restriction to $M^{10}$ of the metric on $M^{11}$.)  
In \ymlag, $\lambda$ is the gauge coupling constant.  We will ultimately 
see that $\lambda\sim \kappa^{2/3}$, so that $L_{YM}$ is of order 
$\kappa^{2/3}$ relative to $L_S$.  
The supersymmetry transformation laws are 
\eqn\ymtrans{\eqalign{\delta A_A^a & = {1\over 2}\bar\eta\Gamma_A\chi^a \cr
                      \delta \chi^a & = -{1\over 4}\Gamma^{AB}F_{AB}^a\eta.
\cr}}
We have here made explicit an index $a=1,\dots,248$ labeling the adjoint 
representation of $E_8$.  We define an inner product by $X^aX^a=\tr\, X^2 
=(1/30)\Tr X^2$, with $\Tr$ the trace in the adjoint representation.  

We wish to add additional interactions to the above and modify the 
supersymmetry transformation laws so that $L_S+L_{YM}+\dots$ will 
be locally supersymmetric.  The first steps are as follows.  Let $T_{YM}$ and 
${\cal S}_{YM}$ be the energy-momentum tensor and supercurrent of the 
supergauge multiplet.  As in any coupling of matter to supergravity, the 
variation of $L_{YM}$ under local supersymmetry contains terms $D_A\bar\eta
{\cal S}_{YM}^A $, reflecting the fact that $L_{YM}$ is only invariant under 
\ymtrans\ if $\eta$ is covariantly constant, and $\bar\eta \psi T_{YM}$, 
coming from the variation of $L_{YM}$ under a local supersymmetry 
transformation of the metric.  To cancel these variations, it is necessary 
-- as usual in supergravity -- to add an interaction 
$\bar\psi{\cal S}_{YM}$.  In the case at hand, this interaction is
\eqn\mummy{L_1=-{1\over 4\lambda^2}\int_{M^{10}}d^{10}x\sqrt g \,\,\bar\psi_A
\Gamma^{BC}\Gamma^AF_{BC}^a\chi^a.}
A small calculation shows that the variation of $L_1$ under local 
supersymmetry cancels the $D\bar \eta \chi F$ and $\eta \psi \chi D\chi$ 
variations of $L_{YM}$, and also cancels part of the $\eta\psi F^2$ 
variation.  The uncanceled variation turns out, after some  gamma matrix 
gymnastics, to be 
\eqn\plummy{\Delta={1\over 16\lambda^2} \int_{M^{10}} d^{10}x\sqrt g\,
\,\bar\psi_A\Gamma^{ABCDE}F^a_{BC}F^a_{DE}\eta.}

Rather as in the coupling of the ten-dimensional vector multiplet 
to {\it ten}-dimensional supergravity \van, there is no way to cancel 
this variation by adding to the Lagrangian additional matter couplings.  
A peculiar mixing of the supergravity and matter multiplets is needed.  

When one verifies the local supersymmetry of the eleven-dimensional 
supergravity Lagrangian $L_S$, it is necessary among other things 
to check the cancellation of the $\eta G D\psi$ and $D\eta G \psi$ 
terms.  In this verification, it is necessary to integrate by parts 
and use the Bianchi identity $dG=0$.%
\foot{This occurs when one varies the interaction $\bar\psi_I\Gamma^{IJKLMN}
\psi_N G_{JKLM}$ with $\delta \bar \psi_I\sim D_I\bar \eta$.  To cancel other 
variations, one must integrate by parts so that the $D_I$ acts on $\psi_N$ 
instead of on $\eta$.  The integration by parts gives a term proportional 
to $dG_{IJKLM}$.}
To cancel $\Delta$, one must modify the Bianchi identity to read 
\eqn\modbi{dG_{11\,ABCD} = - 3\sqrt 2 {\kappa^2\over\lambda^2}\delta(x^{11})
  F^a_{[AB}F^a_{CD]}.}
This correction to the Bianchi identity adds an extra variation of $L_S$
that precisely cancels $\Delta$.

Much as in the analogous story in ten dimensions, \modbi\ implies that the 
three-form $C$ is not invariant under Yang-Mills gauge transformations.  To 
determine the gauge transformation law of $C$, it is convenient to solve the 
modified Bianchi identity by introducing 
\eqn\odbi{\omega_{BCD}= \tr\left(A_B(\partial_CA_D-\partial_DA_C)
+{2\over 3}A_B[A_C,A_D]+{\rm cyclic~permutations~of}\,\,B,C,D\right).}
Thus
\eqn\wodbi{\partial_C\omega_{BCD}+{\rm cyclic~permutations}=6\;\tr\,F_{[AB}
F_{CD]}.}
The Bianchi identity can then be solved by modifying the definition of 
$G_{11\,ABC }$, the new definition being 
\eqn\jodbi{G_{11\,ABC}=\left(\partial_{11}C_{ABC}\pm 23 \,\,{\rm 
permutations}\right)
+{\kappa^2\over \sqrt 2 \lambda^2}\delta(x^{11})\omega_{ABC}.}
Under an infinitesimal gauge transformation $\delta A_A^a=-D_A\epsilon^a$, 
$\omega$ transforms by 
\eqn\rodbi{\delta \omega_{ABC} =\partial_A\left(\tr \,\,\epsilon F_{BC}\right)
+{\rm cyclic~permutations~of}\,\,A,B,C,}
so gauge invariance of $G_{11\,ABC}$ holds precisely if the three-form
$C$ transforms under gauge transformations by
\eqn\hodbi{\delta C_{11\,AB}=-{\kappa^2\over 6\sqrt 2 \lambda^2}\delta(x^{11})
\,\,\tr \,\,\epsilon F_{AB}.}

A correction to the supersymmetry transformation law of $C_{11\,BC}$ is 
also necessary.  It can be determined by requiring that the supersymmetry 
variation of $G_{11\,ABC}$ be gauge-invariant (otherwise this variation 
gives gauge non-invariant, uncancellable terms in the supersymmetry 
variation of the Lagrangian) and is 
\eqn\jodbi{\tilde \delta C_{11\, BC} =-{\kappa^2\over 6\sqrt 2 \lambda^2}
\tr \left(A_B\delta A_C-A_C\delta A_B\right),}
where on the right $\delta A$ is the standard supergravity transformation 
law given in \ymtrans.  With this correction to $\delta C$, the correction 
to $\delta G$ is 
\eqn\codbi{\tilde\delta G_{11\, ABC}={\kappa^2\over \sqrt 2\lambda^2}
\delta(x^{11})\left( \bar\eta\Gamma_A\chi^a F^a_{BC}+{\rm cyclic~permutations
~of}\,\,A,B,C\right).}

\bigskip\noindent
{\it Boundary Behavior}

There is another sense in which we can ``solve the Bianchi identity.''  
We can ask, compatibly with the equation of motion, what can be 
the behavior near $x^{11}=0$ of a $G$ field that obeys the corrected 
Bianchi identity found above, which was 
\eqn\omodbi{dG_{11\,ABCD} = - 3\sqrt 2 {\kappa^2\over\lambda^2}\delta(x^{11})
  F^a_{[AB}F^a_{CD]}.}
How can $dG$ acquire such a delta function?  $G$ itself cannot have a delta 
function at $x^{11}=0$, as that would not be compatible with the equations 
of motion.  However, as $G$ is odd under $x^{11}\to -x^{11}$, it is natural 
for $G_{ABCD}$ to have a step function discontinuity at $x^{11}=0$, giving 
a delta function in $dG$.  In fact, $G_{ABCD}$ must have a jump at $x^{11}=0$ 
given precisely by 
\eqn\omoodbi{G_{ABCD} = - {3 \over \sqrt 2}
 {\kappa^2\over\lambda^2}\epsilon(x^{11}) F^a_{[AB}F^a_{CD]}+\dots.}
Here $\epsilon(x^{11})$ is 1 for $x^{11} >0$ and $-1$ for $x^{11}<0$; the 
$\dots$ are terms that are regular near $x^{11}=0$ and therefore (since $G$ 
is odd under $x^{11}\to -x^{11}$) vanish at $x^{11}=0$.  This is the behavior 
required by the modified equations of motion and Bianchi identity.

This discontinuity means that $G_{ABCD}$ does not have a well-defined 
limiting value as $x^{11}\to 0$.  However, $G^2$ has such a limit, 
which moreover is determined by \omoodbi\ in terms of the gauge fields at 
$x^{11}=0$.  

There is another interesting way to think about \omoodbi.  In this 
paper we are working ``upstairs'' on a smooth eleven-manifold 
$M^{11}$, and requiring $\Z_2$ invariance.  It is natural conceptually 
(though sometimes less convenient computationally) to work 
``downstairs'' on the manifold-with-boundary $X=M^{11}/\Z_2$.  
In that case, it is not natural to add a correction to $dG$ supported 
at the boundary of $X$ (that is, at $x^{11}=0$).  More natural is to 
impose a boundary condition that has the same effect.  Assuming 
that one identifies $X$ with the portion of $M^{11}$ with $x^{11}>0$, 
the requisite boundary condition is simply 
\eqn\jodbi{\left. G_{ABCD}\right|_{x^{11}=0} = - {3  \over \sqrt 2}
 {\kappa^2\over\lambda^2}  F^a_{[AB}F^a_{CD]}.}
(If one identifies $X$ with the $x^{11}<0$ portion of $M^{11}$, one would 
want the opposite sign in \jodbi.)  The idea here is that, since $dG=0$, 
the integration by parts explained in the footnote just before \modbi\ no 
longer picks up a delta function term, but (since there is now a boundary) 
it does pick up a boundary term that has the same effect.%
\foot{In working on $X=M^{11}/\Z_2$ instead of $M^{11}$, one should replace 
the $1/\kappa^2$ in \suglag\ by $2/\kappa^2$, because one is integrating 
\suglag\ over a space of half the volume.  This factor of 2 goes into 
verifying the normalization of \jodbi.}

Thus, in working downstairs on $X$, $G$ has a well-defined boundary value 
given by \jodbi\ (or the same expression with opposite sign if one picks 
orientations opppositely).  In working upstairs on $M$, $G$ does not quite 
have a well-defined value at $x^{11}=0$, but $ G^2$ does.  

\newsec{Analysis of Anomalies}

The most important conclusions of the last section are the gauge 
transformation law \hodbi\ for the three-form $C$, and the formula \omoodbi\ 
for the behavior of $G$ near $M^{10}$.  We will now put these together to get 
an eleven-dimensional view of gauge and gravitational anomalies.  

The idea is that \hodbi\ is analogous to the gauge transformation law 
$\delta B\sim \tr\,\epsilon F$ for the two-form $B$ of string theory, and 
\omoodbi\ will turn the ``Chern-Simons interaction'' $\int C\wedge G\wedge G$ 
of eleven-dimensional supergravity into a Green-Schwarz term.  

We recall that the $CGG$ interaction is, to be precise, a term 
\eqn\plumbo{W=-{\sqrt 2\over 3456\kappa^2}\int_{M^{11}}
\epsilon^{M_1M_2\dots M_{11}} C_{M_1M_2M_3}G_{M_4\dots M_7}
G_{M_8\dots M_{11}}     .}
The variation of $W$ under an arbitrary variation of $C$ is therefore 
\eqn\lumbo{\delta W=-{\sqrt 2\over 1152 \kappa^2}\int_{M^{11}}
\epsilon^{M_1M_2\dots M_{11}} \delta
C_{M_1M_2M_3}G_{M_4\dots M_7}G_{M_8\dots M_{11}}.}
Given that $C$ is not invariant under gauge transformations, 
neither is $W$.  Using \hodbi\ for the gauge variation of $C$, we get 
for the gauge variation of $W$ 
\eqn\jumbo{\delta W=-{1\over 2304 \lambda^2} \int_{M^{10}}
\epsilon^{M_1M_2\dots M_{10}}\epsilon^a F_{M_1M_2}^aG_{M_3\dots M_6}
G_{M_7\dots M_{10}}.}
To proceed further, we need the value of $G^2$ at $x^{11}=0$.  This is given 
by \omoodbi\ in the orbifold approach or equivalently by the boundary 
condition \jodbi\ if one works ``downstairs.''  Either way, one gets 
\eqn\numbo{\delta W = -{\kappa^4\over 128\lambda^6} \int_{M^{10}}
\epsilon^{M_1M_2\dots M_{10}}\epsilon^aF^a_{M_1M_2} F^b_{M_3M_4}F^b_{M_5M_6}
F^c_{M_7M_8}F^c_{M_9M_{10}}.}

So, as promised, the classical theory is not gauge invariant.  There 
is no way to cure this at the classical level.  The only recourse is 
to quantum anomalies.  The anomalous variation of the effective action 
$\Gamma$ for ten-dimensional Majorana-Weyl fermions in an arbitrary 
representation of a simple gauge group is 
\eqn\rumbo{\delta \Gamma={1\over 2}{1\over (4\pi)^55!} 
\int_{M^{10}}\epsilon^{M_1M_2\dots M_{10}} \Tr\left( \epsilon F_{M_1M_2}
F_{M_3M_4} \dots F_{M_9M_{10}}\right),}
with $\Tr$ being the trace in the fermion representation.  The case that we 
are interested in is that the gauge group is $E_8$ and the fermions are in 
the adjoint representation.  In that case, one has the wonderful and unique 
(to $E_8$) identity $\Tr W^6=(\Tr W^2)^3/7200$ (and likewise $\Tr \epsilon 
F^5= \Tr \epsilon F (\Tr F^2)^2/7200$).  If furthermore we write, as is 
customary, $\tr\,W^2=\Tr W^2/30$, then $\Tr W^6=(15/4) (\tr\, W^2 )^3$.  In 
this case, therefore, the quantum anomaly \rumbo\ can be written 
\eqn\urumbo{\delta \Gamma={15\over 8(4\pi)^5 5!} \int_{M^{10}}
\epsilon^{M_1M_2\dots M_{10}}
\,\tr\,(\epsilon F_{M_1M_2})\,\tr\,(F_{M_3M_4}F_{M_5M_6})\,\tr\,(F_{M_7M_8}
F_{M_9M_{10}}).}
It therefore has the right structure to cancel \numbo\ (recall that the 
metric on the Lie algebra was defined by $\epsilon^aF^a=\tr \,\epsilon F$).

Implementing this cancellation, we learn finally that, as promised 
in the introduction, the gauge coupling is related to the gravitational 
coupling by 
\eqn\relcoup{\lambda^2=2\pi\left(4\pi\kappa^2\right)^{2/3}.}

One might have expected that the analogs of the Green-Schwarz terms 
in the present discussion would be boundary interactions, that is 
interactions supported at $x^{11}=0$.  This is not the case, as we have 
seen.  In fact, given a gauge variation of $C$  proportional 
to $\delta(x^{11})$, the possible resulting  gauge variation of a boundary 
interaction would necessarily be proportional to $\delta(0)$.  Thus, 
the ``Green-Schwarz terms'' must be bulk interactions; this goes for 
the ``Chern-Simons''  $CGG$ term  and other terms discussed below.  

\subsec{Extension to Gravitational Anomalies}

We determined the gauge coupling by canceling the purely gauge anomalies
at the boundary of the eleven-dimensional world.  We would now like
to include also the gravitational and mixed anomalies.  

{}From the above discussion,  the anomaly four-form $\hat I_4$ is the 
four-form that appears (multiplied by $\delta(x^{11}) dx^{11}$) in the 
Bianchi identity for $G$.  In our work so far, we have seen only a $\tr\,F^2$ 
term in $\hat I_4$, but in view of the known form of the ten-dimensional 
anomalies, the actual structure must be $\tr\,F^2-(1/2)\tr\,R^2$.  Thus, the 
modified Bianchi identity \modbi\ should be replaced by 
\eqn\nomodbi{dG_{11\,ABCD} = - 3\sqrt 2 {\kappa^2\over\lambda^2}\delta(x^{11})
  \left(F^a_{[AB}F^a_{CD]}-{1\over 2}\tr\,R_{[AB}R_{CD]}\right),}
and the formula \omoodbi\ for the behavior near $x^{11}=0$ should 
correspondingly be replaced by 
\eqn\nomoodbi{G_{ABCD} = - {3 \over \sqrt 2}{\kappa^2\over\lambda^2}
\epsilon(x^{11})\left( F^a_{[AB}F^a_{CD]}-{1\over 2}\tr\,R_{[AB}R_{CD]}+
\dots\right).}
There is also a corresponding local Lorentz transformation law of \nomoodbi, 
analogous to the $E_8$ gauge transformation law \hodbi.  

The $\tr\,R^2$ terms in these formulas are not required by the low energy 
supergravity, but (since they are needed for anomaly cancellation, given 
the structure of the one-loop chiral anomalies), they must be present 
in the full $M$-theory.  The situation is presumably analogous to what 
is seen for the perturbative heterotic string, where the $\tr\,R^2$ terms 
in the analogous formulas arise as corrections of order $\alpha'$.  Note 
that the $\tr\,R^2$ correction will appear in \nomodbi\ and \nomoodbi\ 
with the same coefficient, since \nomoodbi\ is deduced from \nomodbi\ 
by reasoning that was explained above.  

Having understood how $\hat I_4$ enters in eleven dimensions, we would 
like now to understand the origin of $\hat I_8$, or equivalently  to 
complete our understanding of the Green-Schwarz terms.  We have already 
found one of the Green-Schwarz terms above -- the long-familiar 
``Chern-Simons'' interaction of eleven-dimensional supergravity.  This 
particular interaction gives a contribution to $\hat I_8$ that is a multiple 
of $\hat I_4^2$, since the boundary behavior of $G$ is $G\sim \hat I_4$, as 
we have seen.  The other Green-Schwarz terms will have to be bulk 
interactions, as explained at the end of the last subsection, and more 
precisely  will have to be interactions of the form 
\eqn\popeye{{\cal I} =\int_{M^{11}} C\wedge\left(a\,\tr\,R^4+b (\tr\,R^2)^2
\right),}
these being the terms that  have the right sort of gauge and local Lorentz 
variations to cancel chiral anomalies.  Note that it is impossible to add 
to \popeye\ terms that  directly involve $F$, since the gauge fields 
propagate only on $M^{10}$.  It is also impossible for $F$-dependence to 
arise indirectly from the behavior of $G$ near $M^{10}$, since \popeye\ is 
independent of $G$; the $C\wedge G\wedge G $ has already been taken into 
account (with a coefficient known from low energy supergravity), and a term 
$C\wedge G\wedge\tr\,R^2$ is not possible, as it would violate the parity 
symmetry of $M$-theory.

${\cal I}$ will contribute to $\hat I_8$ a term that involves $R$ only, so we 
get the striking prediction that $\hat I_8$ is a multiple of $\hat I_4^2$ 
plus an eight-form constructed only from $R$.  At this point, it is helpful 
to note that 
\eqn\ploop{\hat I_8 = -{1\over 4}\left(\tr\,F^2-{1\over 2}\tr\,R^2\right)^2
+\left(-{1\over 8}\tr\,R^4+{1\over 32} (\tr\,R^2)^2\right),}
and thus has the expected form.  This is a satisfying test of $M$-theory, as 
this structure of $\hat I_8$ has no known rationale in perturbative string 
theory.  

Actually, we can be more precise, since the interaction \popeye\ is known (at 
least up to an overall multiplicative constant; fixing this constant requires 
a more precise comparison of the normalizations of string theory and 
$M$-theory or a precise knowledge of the two-brane and five-brane tensions 
in $M$-theory).  As we explained in the introduction, the interaction 
\popeye\ is known up to a constant multiple either from comparison to a 
one-loop calculation for Type IIA superstrings \vafa\ or from anomaly 
cancellation for eleven-dimensional five-branes \refs{\duffliu - \witten}.  
Either way, one finds that \popeye\ is a multiple of 
\eqn\guffy{\frac{\sqrt 2}{(4\pi)^3(4\pi\kappa^2)^{1/3}}\int_{M^{11}}
C\wedge\left(-{1\over 8}\tr\,R^4 +{1\over 32}(\tr\,R^2)^2\right).}
Thus, at least the relative coefficient of $\tr\,R^4 $ and $(\tr\,R^2)^2$ 
agrees with the ``experimental'' structure of $\hat I_8$.  This is again a 
real test of $M$-theory, since there is no perturbative string theory reason 
for this to work.  The structure of \popeye\ is deduced either via Type IIA 
perturbation theory or anomaly cancellation for eleven-dimensional 
five-branes, and any known rationale for comparing the anomaly polynomial of 
the perturbative heterotic string to either of these involves $M$-theory.  

Notice that the coefficient $-1/4$ of the $\hat I_4^2$ term in $\hat I_8$ 
is a matter of convention; it could be shifted by a scaling $\hat I_4\to u 
\hat I_4$, $\hat I_8\to u^{-1} \hat I_8$, without affecting the 
factorization $\hat I_{12}=\hat I_4\hat I_8$.  Modulo this imprecision 
in the definition of $\hat I_8$, we have from $M$-theory a complete 
{\it a priori\/} prediction for $\hat I_8$, which amounts to a prediction for
three numbers ($\hat I_8$ is a linear combination of four monomials 
$(\tr\,F^2)^2$, $\tr\,F^2\,\tr\,R^2$, $(\tr\,R^2)^2$, and $\tr\,R^4$, but 
the coefficient of one monomial can be scaled out as just explained).  
There are therefore three predictions, of which we have here verified two; 
verification of the last prediction requires a more precise comparison of 
different conventions, as noted in the last paragraph.  

\newsec{Construction of the Lagrangian}

In this section, we will proceed with additional steps in the construction
of the locally supersymmetric Lagrangian.  The formula $\lambda^2=2\pi(4\pi
\kappa^2)^{2/3}$ obtained in the last section is of some conceptual interest 
in organizing the computation.  It shows that the theory has only one natural 
length scale, given by $\kappa^{2/9}$.  Moreover, on dimensional grounds, 
the decomposition of the boundary interactions in terms with more and more 
matter fields is an expansion in powers of $\kappa$.  The leading boundary 
interactions (the minimal Lagrangian $L_{YM}$ of the gauge multiplet 
and terms related to it by supersymmetry) are of order $\kappa^{2/3}$ 
relative to the gravitational action.  Formally, the construction of the 
locally supersymmetric classical action appears to be an expansion in 
integral powers of $\kappa^{2/3}$. Other exponents must arise in the actual 
quantum $M$-theory, since we will run into infinities which, when cut off in 
the quantum theory, must on dimensional grounds give anomalous powers of 
$\kappa$.  

There are two principal goals of the rather complicated computation performed 
in this section:  

(1) To add to our confidence that the supersymmetric coupling of 
the vector multiplet on $M^{10}$ to the supergravity multiplet on $M^{11}$ 
does exist, by working out the classical construction of this coupling 
to the extent that it makes sense.  

(2) To exhibit the limits of the classical construction (beyond what 
is evident from the discussion of anomalies in section three) by 
showing how infinities arise in order $\kappa^{4/3}$.  

In the computation, one can be guided to a certain extent by the 
{\it ten}-dimensional coupling of the vector and supergravity multiplets 
\refs{\van,\chap}, to which our discussion must reduce at low energies in the 
appropriate limit.  This gives clues to many of the terms that must be added 
to the Lagrangian and transformation laws.  On the other hand, in the 
computation one definitely meets terms (involving $D_{11}\eta$, for 
instance) that vanish upon dimensional reduction to ten dimensions but must 
be canceled to achieve local supersymmetry in eleven dimensions.  Thus, the 
existence of the coupling we are constructed (and again, we believe that it 
only exists in full at the quantum level) goes well beyond ten-dimensional 
considerations.  

We will carry out the computation in three stages: (i) first we 
complete the construction of the Lagrangian in order $\kappa^{2/3}$; 
(ii) then we look at some terms in order $\kappa^{4/3}$; (iii) finally 
we look systematically at all four-fermi variations in order $\kappa^{2/3}$.  
It might seem illogical to put (ii) before (iii).  We have done this 
because (ii) is much simpler than (iii), and also gives an easy way 
to determine some of the transformation laws that are needed in (iii).  

Of course, we cannot hope for a full determination of the structure.  
Apart from requiring a much fuller knowledge of the quantum mechanics of 
$M$-theory than one has, the full structure is presumably non-polynomial, 
like the $\alpha'$ expansion of perturbative string theory.  Once one reaches 
a sufficiently high order in $\kappa$, one would require among other things 
a more complete knowledge of the low energy expansion of $M$-theory in bulk 
(including higher derivative interactions) in order to proceed.  

\subsec{Some New Interactions}

The boundary interactions (that is, interactions supported at $x^{11}=0$) 
that we discussed in section two are the minimal super Yang-Mills action 
and the supercurrent coupling 
\eqn\nogo{L_0=-{1\over 2\pi (4\pi\kappa^2)^{2/3}}\int_{M^{10}}d^{10}x\sqrt g
\,\left({1\over 4}\,\tr\,F_{AB}F^{AB}+{1\over 2}\,\tr\,\bar\chi\Gamma^AD_A
\chi+{1\over 4} \bar\psi_A\Gamma^{BC}\Gamma^AF_{BC}^a\chi^a\right).}
These terms are all of order $\kappa^{2/3}$ compared to the supergravity 
action.  There is precisely one more boundary interaction of the same order.  
To find it, one can look at the terms of order $F\chi G\eta$ in the 
supersymmetry variation of the Lagrangian.   One source of such terms comes 
from the variation of the supercurrent interaction in \nogo\ with 
$\delta\psi \sim G\eta$.  Another source comes as follows.  
We found in section two a correction \codbi\ to the supersymmetry variation 
of $G_{ABC\,11}$.  The $G_{ABC\,11}^2$ term in the bulk supergravity action 
therefore picks up a new variation supported at $x^{11}=0$; this term is 
again proportional to $F\chi G\eta$.  These terms by themselves do not 
cancel.  After a moderately lengthy computation, one finds that to cancel 
them one must add a new boundary interaction, 
\eqn\plummo{L_1={\sqrt 2\over 96\pi(4\pi\kappa^2)^{2/3}}\int_{M^{10}}d^{10}x
\sqrt g\,\,\bar\chi^a\Gamma^{ABC}\chi^aG_{ABC\,11}.}
This term -- and the verification that the $F\chi G\eta$ terms cancel -- is 
quite similar to an analogous term and verification in the ten-dimensional 
supergravity/Yang-Mills coupling.  

This actually completes the construction of the Lagrangian in order 
$\kappa^{2/3}$ and verification of local supersymmetry up to four-fermi 
terms, whose analysis we postpone to the next subsection.  
Instead we turn to something that is of conceptual interest and 
still relatively simple.  

In \plummo, we see an interaction in which $G_{ABC\,11}$ is evaluated on 
$M^{10}$, that is at $x^{11}=0$.  On the other hand, in \codbi, we found a 
term in the supersymmetry variation of $G_{ABC\,11}$ that is proportional to 
$\delta(x^{11})$.  If we combine the two, that is if we vary $L_1$ according 
to \codbi, we get a result proportional to $\delta(0)$.  
This presumably should be interpreted as a linear divergence that is cut off 
somehow in the quantum $M$-theory.  For our present purposes, though, we will 
be pragmatic, and without worrying about precisely what $\delta(0)$ means, we 
will attempt to formally cancel the $\delta(0)$ terms.  

Obviously, to do this we need more sources of $\delta(0)$ terms.  
Since the term we want to cancel is proportional to $\chi\chi\chi F\eta$, 
there are two sources of terms that might cancel it.  We could 
add to the gravitino variation an extra term $\delta\psi_A\sim\delta(x^{11}) 
\chi\chi\eta$.  When combined with the $\chi F\psi$ interaction in 
$L_0$, it gives another term of the general form $\delta(0)\chi^3F\eta$.  
Finally, one could add to the Lagrangian a term $\delta(0)\chi^4$, 
which will again have a variation of the desired form.  After another 
moderately long calculation, one finds that the new terms required in the 
gravitino variation are%
\foot{Here $\delta(x^{11})$ is understood as the delta function that 
transforms as a scalar under diffeomorphisms; this involves an implicit 
power of the 11-11 component of the frame field $e_{11}{}^{11}$, and is 
required to match the transformation properties of $\delta\psi_A$.}
\eqn\gravar{\delta\psi_A=-{1\over 576\pi}\left(\kappa\over 4\pi\right)^{2/3} 
\delta(x^{11})\;\bar\chi^a\Gamma_{BCD}\chi^a\left(\Gamma_A{}^{BCD}\eta-
6\delta_A^B\Gamma^{CD}\eta\right),}
and that the new interaction required is
\eqn\hubbu{L_\chi=-{\delta(0)\over 96 (4\pi)^{10/3} \kappa^{2/3}}
\int_{M^{10}}d^{10}x \sqrt g\,\,\bar\chi^a\Gamma^{ABC}\chi^a\,
\bar\chi^b\Gamma_{ABC} \chi^b.}
The $\delta(0)$ presumably means that in the quantum theory this 
interaction is really of order $\kappa^{-8/9}$, that is, of order 
$\kappa^{10/9}$ relative to the original supergravity action.  

We focussed here on a particular four-fermi term of relative 
order $\kappa^{4/3}$ because it enabled us to exhibit in a simple 
fashion the ``divergences'' that appear in trying to construct 
the classical Lagrangian.  In the next section, we look systematically 
at the four-fermi terms of order $\kappa^{2/3}$.  

\subsec{Four-Fermi Terms in Order $\kappa^{2/3}$} 

To this order, the structure of the Lagrangian and the supersymmetry 
variation of the fields can be determined by canceling terms ${}\sim\chi\chi
\psi\eta$ with one covariant derivative acting on one of the fermi fields.  
There are two natural classes of such terms, depending on whether the 
derivative is normal to the boundary, or acts along the boundary.  

First consider the class of terms containing the normal derivative, i.e.\ 
terms proportional to $D_{11}\eta$ or $D_{11}\psi_A$.  Such terms clearly 
vanish upon the dimensional reduction to ten dimensions: {}from the point of 
view of a ten-dimensional low-energy observer, they only contain 
contributions from massive Kaluza-Klein modes of $\psi_A$ and $\eta$ that 
decouple from the low-energy modes as the radius of the eleventh dimension 
goes to zero.  In the present case, however, the cancellation of such terms 
is not automatic, and will help us determine some new additions to the 
Lagrangian.  

What are the possible sources of terms proportional to $D_{11}\psi_A$ or 
$D_{11}\eta$?   One source of such terms is generated by the correction 
\gravar\ to the variation of $\psi_A$ determined in the previous subsection.  
Since this correction is proportional to the delta function localized at the 
boundary, it will generate boundary four-fermi terms proportional to $D_{11}
\psi_A$ when applied to the bulk kinetic term of the gravitino.  Given 
\gravar , this variation generates just one term, equal to
\eqn\mravenec{-\frac{1}{64\pi(4\pi\kappa^2)^{2/3}}\int_{M^{10}}d^{10}x\sqrt g
\,\,\bar\chi^a\Gamma^{ABC}\chi^a\;\bar\eta\Gamma_{AB}D_{11}\psi_C.}
This term cancels exactly against a similar term that comes from the bulk 
variation of $G_{ABC\,11}$ in the interaction term $L_1$.  No other terms 
with $D_{11}\psi_A$ appear in the supersymmetry variation of the Lagrangian 
at this order.  

As to the terms with $D_{11}\eta$, they have two sources among the terms 
already present in the Lagrangian.  First of all, the bulk variation of 
$G_{ABC\,11}$ in $L_1$ produces a term proportional to $\bar\chi\Gamma^{ABC}
\chi\;\bar\psi_A\Gamma_{BC}D_{11}\eta$.  This term can only be canceled if 
we introduce a new interaction, 
\eqn\rakosnik{L_2=\frac{1}{64\pi(4\pi\kappa^2)^{2/3}}\int_{M^{10}}d^{10}x
\sqrt g\,\,\bar\chi^a\Gamma^{ABC}\chi^a\,\,\bar\psi_A\Gamma_{BC}\psi_{11}.}
(When $L_2$ is varied, $\delta\psi_A\sim D_A\eta$ leads to a 
$\chi\chi\psi_{11}D_A\eta$ term; this term cancels against the term of the 
same form that comes from the variation of $G_{ABC\,11}$ in $L_1$.  In 
addition, the variations of the bulk supergravity action
$L_S$ and of $L_1$ give terms of the form 
$\chi\chi\eta D_A\psi_{11}$; happily, these  terms cancel against each 
other.)  

Another source of terms proportional to $D_{11}\eta$ is the variation of the 
spin connection in the gaugino kinetic term.  This point requires a further 
explanation.  In our treatment of eleven-dimensional supergravity in bulk, 
we have adopted the 
1.5 order formalism, which means that the spin connection is 
first treated as an independent variable and 
set equal to the solution of its bulk equations of motion at the end of the 
calculation.  
One then need not worry about the supersymmetry variation of the spin
connection, which vanishes by the equations of motion.
In including the boundary interactions, we
prefer to continue to use the ``bulk'' formula for the spin connection. 
This means that when the spin connection appears in boundary interactions,
its supersymmetry variation must be included.  (One could avoid this
by extending the 1.5 order formalism to incorporate boundary corrections
to the spin connection determined by the equations of motion, but we did
not find that approach simpler.) 

In practice, to the order we will calculate, the spin connection only appears 
in 
the gravitino kinetic term.  Its variation produces an  additional term 
proportional to $D_{11}\eta$ (plus other terms we will consider
later). Canceling this $D_{11}\eta$ term requires a new term in the 
Lagrangian, 
\eqn\rumcajs{L_3=\frac{1}{64\pi(4\pi\kappa^2)^{2/3}}\int_{M^{10}}d^{10}x
\sqrt g\,\,\bar\chi^a\Gamma_{ABC}\chi^a\,\,\bar\psi_D\Gamma^{DABC}\psi_{11}.}

Now we will show that no other $\psi_{11}$ dependent four-fermi terms 
are generated in the Lagrangian at this order in $\kappa$, beyond those given 
by \rakosnik\ and \rumcajs .   To see this, we will proceed as follows.  
Supersymmetry variation of such additional four-fermi terms would produce 
additional terms proportional to $D_{11}\eta$.  Notice that these 
$D_{11}\eta$ terms could only be canceled if there is a three-fermi 
correction to the variation of the gravitino, $\delta'\chi\sim\chi\psi_{11}
\eta$, and the gaugino kinetic term is varied.  We will prove our claim that 
no new $\psi_{11}$-dependent four-fermi terms arise in the Lagrangian at this 
order, by showing that there are no $\chi\psi_{11}\eta$ corrections to the 
supersymmetry variation of $\chi$.  Upon variation of the gaugino kinetic 
term, such corrections would produce terms of the form 
\eqn\kremilek{\bar\chi\Gamma_A D_B\chi\;\bar\psi_{11}\ldots\eta\quad{\rm and}
\quad\bar\chi\Gamma_{ABCDE}D_F\chi\,\bar\psi_{11}\ldots\eta.}
(Here $\ldots$ denotes all possible combinations of $\Gamma$ matrices.)  
There is no other possible source of such terms; a simple calculation shows 
that their cancellation requires the supersymmetry variation of $\chi^a$ to 
be independent of $\psi_{11}$, thus completing our argument.  

Having canceled all terms with $D_{11}\psi$ and $D_{11}\eta$, we can 
determine the rest of the structure at this order in $\kappa$ by looking at 
cancellations of $\chi\chi\eta\psi$ terms where now the {\it ten}-dimensional 
derivative $D_A$ acts on one of the four fermions.  First we determine the 
correction to the supersymmetry variation of $\chi^a$, by canceling terms of 
the form 
\eqn\vochomurka{\bar\chi\Gamma_A D_B\chi\;\bar\eta\ldots\psi_C\quad{\rm and}
\quad\bar\chi\Gamma_{ABCDE}D_F\chi\;\bar\eta\ldots\psi_G.}
Terms of this structure must cancel by themselves, since chirality and fermi 
statistics do not allow one to use integration by parts to move the 
derivative away from the gauginos.  There are two obvious sources of such 
terms: the variation of $e_A{}^m$ in the gaugino kinetic term, and the 
variation of $F_{AB}^a$ in the supercurrent coupling of $L_0$.  As these do 
not cancel, one has to look for another source of such terms.  We can add 
a correction, $\delta'\chi\sim\chi\psi\eta$, to the supersymmetry variation 
of the gauginos.  This correction will produce terms of the required form 
\vochomurka\ from the variation of the gaugino kinetic term, and the precise 
form of $\delta'\chi$ will be determined from the cancellation of these 
terms.%
\foot{{}In this and some of the following calculations, we need a Fierz 
rearrangement formula for chiral ten-dimensional fermions.  All rules follow 
from the expansion of the product of two fermions $\xi$ and $\zeta$ on 
$M^{10}$ that obey $\Gamma_{11}\xi=\xi$ and $\Gamma_{11}\zeta=\zeta$: 
$$\zeta^\alpha\bar\xi_\beta=-\frac{1}{32}\left(2\left(\bar\xi\Gamma_A\zeta
\right)\Gamma^{A\,\alpha}{}_\beta-\frac{1}{3}\left(\bar\xi\Gamma_{ABC}\zeta
\right)\Gamma^{ABC\,\alpha}{}_\beta+\frac{1}{120}\left(\bar\xi\Gamma_{ABCDE}
\zeta\right)\Gamma^{ABCDE\,\alpha}{}_\beta\right).$$}
After a tedious calculation, one obtains 
\eqn\sebestova{\eqalign{\delta'\chi^a&=\frac{1}{64}\left(7\left(\bar\psi_A
\Gamma_B\eta\right)\Gamma^{AB}\chi^a+9\left(\bar\psi_A\Gamma^A\eta\right)
\chi^a-\frac{1}{2}\left(\bar\psi_A\Gamma_{BCD}\eta\right)\Gamma^{ABCD}\chi^a
\right.\cr
&\qquad\left.{}-\frac{5}{2}\left(\bar\psi^A\Gamma_{ABC}\eta\right)\Gamma^{BC}
\chi^a+\frac{1}{24}\left(\bar\psi^A\Gamma_{ABCDE}\eta\right)\Gamma^{BCDE}
\chi^a\right).\cr}}

The correction \sebestova\ to the supersymmetry variation of $\chi^a$ can be 
simplified considerably by the Fierz rearrangement formula, leading to 
\eqn\mach{\delta'\chi^a=\frac{1}{4}\left(\bar\psi_A\Gamma_B\chi^a\right)
\Gamma^{AB}\eta.}
This is exactly what one would have expected from the requirement that the 
total supersymmetry transformation of $\chi^a$ be ``supercovariant.''  This 
also explains why no $\psi_{11}$-dependent corrections to the supersymmetry 
variation of $\chi^a$ arise -- when varied, such terms would produce terms 
with $D_{11}\eta$, and supercovariance of the total supersymmetry variation 
of the gauginos would be spoiled.  

Given the correction \sebestova\ to the supersymmetry variation of the 
gauginos, the terms that remain to be determined at this order in $\kappa$ 
are: 

(1) The correction to the supersymmetry variation of $\psi_{11}$; on the 
basis of chirality and fermi statistics, this correction can only be 
proportional to $\left(\bar\chi^a\Gamma_{ABC}\chi^a\right)\Gamma^{ABC}\eta.$ 

(2) Coefficients of all possible $\chi\chi\psi_A\psi_B$ terms in the 
Lagrangian;  there are exactly four possible inequivalent terms of this 
structure.  We will see momentarily that these additional four-fermi terms 
do appear in the Lagrangian.  

We start by canceling terms ${}\sim\bar\chi\Gamma^{ABC}\chi\;\bar\psi_D
\ldots\eta$ with $D_E$ on one of the four fermions.  If the derivative is on 
one of the gauginos, we can now use integration by parts to move it to either 
$\psi_A$ or $\eta$.  This leaves us with two classes of terms to cancel -- 
one with $D_A\psi_B$, and one with $D_A\eta$.  The $\chi\chi\eta D\psi$ 
terms do not get any contribution from the so far undetermined 
$\chi\chi\psi_A\psi_B$ terms in the Lagrangian, since at this order those 
will only contribute to $\chi\chi D\eta\psi$ terms.  Hence, we can use 
cancellation of the $\chi\chi\eta D\psi$ terms to determine the correction to 
the supersymmetry variation of $\psi_{11}$; another lengthy calculation leads 
to%
\foot{The only subtlety here is related to the cancellation of terms 
$\bar\chi^a\Gamma^{ABC}\chi^a\;\bar\eta\Gamma_AD_B\psi_C$, which  gets a 
contribution from the variation of the spin connection in the gaugino 
kinetic term; recall the discussion of the 1.5 order formalism above.}
\eqn\jonatan{\delta'\psi_{11}=\frac{1}{576\pi}\left(\frac{\kappa}{4\pi}
\right)^{2/3}\delta(x^{11})\left(\bar\chi^a\Gamma^{ABC}\chi^a\right)
\Gamma_{ABC}\eta.}

Once $\delta'\psi_{11}$ has been determined, we can go on and calculate the 
$\bar\chi\Gamma_{ABC}\chi^a\;\bar\psi D\ldots D_E\eta$ terms; their 
cancellation will determine the coefficients of the remaining four-fermi 
terms in the Lagrangian.  (As in the case of the $\chi\chi\eta D\psi$ 
terms, there will be a non-zero contribution from the variation of the spin 
connection in the gaugino kinetic term.)  After some additional algebra, one 
obtains
\eqn\maxipes{\eqalign{L_4&=\frac{1}{256\pi(4\pi\kappa^2)^{2/3}}\int_{M^{10}}
d^{10}x\sqrt g
\,\,\bar\chi^a\Gamma^{ABC}\chi^a\left(3\bar\psi_A\Gamma_B\psi_C-\frac{}{}
\bar\psi_A\Gamma_{BCD}\psi^D\right.\cr
&\left.\qquad\qquad\qquad\qquad{}-\frac{1}{2}\bar\psi_D\Gamma_{ABC}\psi^D
-\frac{13}{6}\bar\psi^D\Gamma_{DABCE}\psi^E\right).}}
This completes the construction of the boundary
Lagrangian to order $\kappa^{2/3}$, 
which is thus equal to the sum $L=L_0+L_1+L_2+L_3+L_4$, with the individual 
terms given by \nogo , \plummo , \rakosnik , \rumcajs\ and \maxipes .  

We could stop our discussion here; instead, however, one simple point seems 
worth making.  It turns out that the four-fermi terms that we found at order 
$\kappa^{2/3}$ are exactly those implied by supercovariance to this order 
in $\kappa$, and can therefore be absorbed into the definition of 
supercovariant objects.  This allows us to summarize the structure of all 
boundary terms in the Lagrangian at order $\kappa^{2/3}$ as constructed in 
this section, in the following succinct formula: 
\eqn\manka{\eqalign{L&=\frac{1}{2\pi(4\pi\kappa^2)^{2/3}}\int_{M^{10}}d^{10}x
\sqrt g\,\left(\frac{1}{4}\,\tr\,F_{AB}F^{AB}+\frac{1}{2}\,\tr\,\bar\chi
\Gamma^AD_A(\hat\Omega)\chi\right.\cr
&\qquad\left.{}+\frac{1}{8}\bar\psi_A\Gamma^{BC}\Gamma^A(F^a_{BC}+
\hat F^a_{BC})\chi^a+\frac{\sqrt 2}{48}\bar\chi^a\Gamma^{ABC}\chi^a\;
\hat G_{ABC\,11}\right).\cr}}
Here the supercovariant spin connection $\hat\Omega_A^{mn}$, Yang-Mills field 
strength $\hat F_{AB}^a$, and field strength $\hat G_{ABC\,11}$ are given by 
\eqn\cipisek{\eqalign{\hat\Omega_{A\,BC}&=\Omega_{A\,BC}+\frac{1}{8}
\bar\psi^D\Gamma_{DABCE}\psi^E-\frac{1}{4}\bar\psi^D\Gamma_{DABC}\psi_{11},\cr
\hat F^a_{AB}&=F^a_{AB}-\bar\psi_{[A}\Gamma_{B]}\chi^a,\cr
\hat G_{ABC\,11}&=G_{ABC\,11}+\frac{3\sqrt 2}{4}\left(\bar\psi_{[A}
\Gamma_{BC]}\psi_{11}-\bar\psi_{[A}\Gamma_B\psi_{C]}\right).}}
(In accord with the version of the 1.5 order formalism used in this paper, 
the spin connection $\Omega_{A\,BC}\equiv e_{Bm}e_{Cn}\Omega_A^{mn}$ in 
\cipisek\ is a composite of $e_A{}^m$ and $\psi_A$, and solves the bulk 
equations of motion.)  

Hence, we see that -- just as in the case of pure eleven-dimensional 
supergravity \julia\ -- no further four-fermi terms are generated at order 
$\kappa^{2/3}$ beyond those required by eleven-dimensional supercovariance.  
Of course, at higher orders in $\kappa$ we encounter additional four-fermi 
terms that are not explained in this way -- the first example of such terms 
is the term $L_\chi$ of \hubbu , which is quartic in the gauginos and appears 
at relative order $\kappa^{4/3}$.  

\listrefs
\end